# Giant sharp magnetoelectric switching in multiferroic epitaxial La$_{0.67}$Sr$_{0.33}$MnO$_3$ on BaTiO$_3$

W. Eerenstein[1*], M. Wiora[1,2*], J.L. Prieto[1,3], J.F. Scott[4] and N.D. Mathur[1]

1. University of Cambridge, Department of Materials Science, Pembroke Street, Cambridge CB2 3QZ, United Kingdom.
2. University of Applied Science Ulm, Faculty of Mechatronics, Albert-Einstein-Allee 55, 89081 Ulm, Germany.
3. Instituto de Sistemas Optoelectronicos y Microelectronica - E.T.S. Ing. Telecomunicaciones - UPM, Avd. Complutense s/n - 28040 – Madrid, Spain.
4. University of Cambridge, Department of Earth Sciences, Downing Street, Cambridge, CB2 3EQ, United Kingdom.

[*]These authors contributed equally to this work.

**Magnetoelectric coupling permits a magnetic order parameter to be addressed electrically or vice versa[1], and could find use in data storage, field sensors and actuators. Coupling constants for single phase materials such as chromium dioxide[2], boracites[3] and manganites[4,5] are typically as low as $10^{-12}$-$10^{-9}$ s m$^{-1}$, e.g. because the polarisations and magnetisations are small. Two phase multiferroics with strain mediated coupling, such as laminates[6], composites[7] and epitaxial nanostructures[8], are more promising because each phase may be independently optimised. The resulting magnetoelectric switching can be larger, e.g. $10^{-8}$ s m$^{-1}$, but it is not sharp because clean coupling is precluded by the complexity of the microstructures and concomitant strain fields. Here we report a giant sharp magnetoelectric effect at a single epitaxial interface between a 40 nm ferromagnetic stress-sensitive La$_{0.67}$Sr$_{0.33}$MnO$_3$ film, and a 0.5 mm BaTiO$_3$ substrate that is ferroelectric, piezoelectric and ferroelastic. By applying a small electric field (4-10 kV cm$^{-1}$) across the entire structure, we achieve persistent changes in film magnetisation of up to 65% near the BaTiO$_3$ structural phase transition at ~200 K. This represents a giant magnetoelectric coupling ($2.3 \times 10^{-7}$ s m$^{-1}$) that arises from strain fields due to ferroelastic non-180° domains whose presence we confirm using x-ray diffraction. The coupling persists over a wide range of temperatures including room temperature, and could therefore inspire a range of sensor and memory applications.**

In a given material or system, the magnetoelectric (ME) coupling of electric and magnetic degrees of freedom opens up new avenues for device design. For example, in data storage one could write information electrically and read it magnetically, thus exploiting the best aspects of ferroelectric random access memory (FeRAM) and magnetic random access memory (MRAM). In other examples[9], sensor and actuator devices could benefit from the use of remote magnetic fields to generate electrical signals and movement respectively. ME coupling is quantified as either a change of magnetization $dM$ per applied electric field change $dE$, or a change of electrical polarization $dP$ per applied magnetic field change $dH$. The corresponding linear ME effects are by thermodynamics of equal strength $\alpha = \mu_0 dM/dE = dP/dH$, but the vast majority of experimental studies measure the electrical response to a magnetic stimulus ($\mu_0$ is the permeability of free space). In the opposite scenario we obtain, in response to a finite change in electric field $\Delta E$, a giant and sharp persistent change $\Delta M$ in a multiferroic heterostructure that comprises an epitaxial ferromagnetic stress-sensitive film on a substrate that is ferroelectric, piezoelectric and ferroelastic.

In single phase materials, $\alpha$ is found to be small. In principle, this is because it is bounded[9,10] by the product of dielectric permittivity and magnetic susceptibility, assuming direct rather than strain mediated coupling. Consequently, large values could arise in multiferroic materials that are both ferromagnetic and ferroelectric. However, ferroic order is not necessarily associated with large response functions[9], multiferroics are rare[11], and higher order ME effects could dominate[9]. In practice, most reports of ME centre on ferroelectrics that are antiferromagnetic. Indeed, ME coupling was discovered[2] in $Cr_2O_3$, where a value of $\alpha=dP/dH=4.1\times10^{-12}$ s m$^{-1}$ was recorded near the Néel temperature of 307 K. At low temperatures, similar strength effects occur in e.g. $Gd_2CuO_4$ [12] and $BiFeO_3$ (where large magnetic fields of 20 T were used to unwind the spiral spin structure)[13], and the effects in $TbMn_2O_5$ [4] and $TbMnO_3$ [5] are only an order of magnitude larger. Also at low temperautes, magnetically induced sharp transitions in the ferroelectric polarization were found in weakly ferromagnetic $Ni_3B_7O_{13}I$ [3] and paramagnetic $Tb_2(MoO_4)_3$ [14], but in both cases $\alpha=\Delta P/\Delta H\sim10^{-9}$ s m$^{-1}$ was limited by the extremely small absolute values of polarization $\ll 1$ μC cm$^{-2}$.

The strength of strain mediated ME coupling is unbounded, and has been widely investigated in two phase material systems that comprise a piezoelectric (or electrostrictive) and magnetostrictive (or piezomagnetic) material. Such heterostructures have been prepared as laminates[6], and composites[7] and self-assembled nanostructures[8] with large area interfaces, but ME coupling tends to be presented[9] as $dE/dH$ rather than $dP/dH$. This is because output voltages are easier to measure than polarizations, especially under quasi-static conditions. To establish values of $\alpha$ for absolute comparison requires knowledge of the relative dielectric permittivity $\varepsilon$. The largest value[6] of $dE/dH$=4800 mV cm$^{-1}$ Oe$^{-1}$ was obtained with laminated $Pb(Zr,Ti)O_3$/terfenol-D, and taking[15] e.g. $\varepsilon$=300 for a typical film of $Pb(Zr,Ti)O_3$ gives $\alpha\sim10^{-8}$ s m$^{-1}$. This represents a quantitative improvement over single phase materials, but the quality of the effect in composites, laminates and nanostructures is compromised because the ME switching is not sharp in view of the microstructural complexity.

ME studies in epitaxial structures are attractive because the homogeneity of the induced strain is maximised even though the interfacial area is not. This improvement arises because the two phases are juxtaposed with a precise crystallographic relationship, and because the interfacial roughness is low. In an experiment by Lee *et al.*[16] where no electric field was applied, the magnetization of an epitaxial ferromagnetic 50 nm $La_{0.67}Sr_{0.33}MnO_3$ (LSMO) film was shown to display discontinuous jumps when the underlying ferroelectric $BaTiO_3$ (BTO) substrate was cooled through structural phase transitions. We expand on this work by investigating the effect of an electric field applied across a similar sample in the variable temperature stage of a Vibrating Sample Magnetometer (VSM). The use of BTO is particularly attractive here because over a wide range of temperatures it displays[17] large ~1% piezoelectric strains associated with ferroelastic non-180° domains. Consequently, electrically addressed BTO has the capacity to alter the structural symmetry and thus the magnetic anisotropy of an overlying epitaxial film.

LSMO is a bad metal and a pseudo-cubic perovskite with $a$=3.87 Å at room temperature[18]. Bulk samples display ferromagnetism below[19] ~365 K, and this Curie temperature is sensitive[18] to epitaxial or grain boundary strain. LSMO films on cubic substrates show biaxial magnetic anisotropy, but similar films on orthorhombic substrates show uniaxial magnetic anisotropy[20]. The perovskite BTO is a well known room temperature ferroelectric that is tetragonal with $a$=$b$=3.99 Å and $c$=4.03 Å [21]. At high temperatures it is cubic (C) and thus not ferroelectric, but on cooling it becomes tetragonal (T) below the Curie temperature of 410 K, orthorhombic (O) below 290 K and rhombohedral (R) below 190 K [21]. The symmetries of the T, O and R states permit non-180° domains such that the ferroelectric phases of BTO are ferroelastic. The largest changes in strain occur through 90° domain switching, and although the mobility of 90° domain walls can be low[22], it has recently been shown[23] that wall mobility is significantly increased if even just a small fraction of other non-180° domains are present.

Five 40 nm LSMO films were grown on 0.5 mm BTO (001) substrates by pulsed laser deposition (Methods). Lee *et al.*[16] used pre-poled BTO in order to minimise cracking associated with thermal excursions to their growth temperature of 750 °C (C.-B. Eom, private communication). By contrast, we used unpoled BTO which is significantly cheaper. Prior to performing ME measurements, we repeated the work of Lee *et al.*[16] by investigating the temperature dependence of film magnetization in the absence of an applied electric field. Data for a typical sample (Sample #1) are shown in Fig. 1. Near the R-O transition, the observed 0.45 $\mu_B$/Mn jump is approximately double the 0.19 $\mu_B$/Mn jump seen by Lee *et al.*[16]. Near the O-T transition, we do not reproduce their sharp jump (0.79 $\mu_B$/Mn) except in one unrepresentative sample where we find a small jump of 0.17 $\mu_B$/Mn (Sample #2). (Magnetization-temperature data for all five samples appear in Supplementary Information.) These differences between the two studies could arise because the microstructures of the BTO substrates are modified by pre-poling, even though BTO is cubic at the 775° C film growth temperature. Alternatively, these differences between the two studies could arise in connection with some degree of self-poling[24], which is a well-known phenomenon arising through flexoelectricity when cooling epitaxial ferroelectric thin films.

X-ray diffraction (XRD) of an as-prepared sample (Sample #1) revealed the expected 90° domains in the room temperature T phase of BTO (Fig. 2a). The broad LSMO film peak corresponds to an out-of-plane lattice parameter (3.86 Å) that closely matches the bulk value (3.87 Å). Given the LSMO-BTO lattice mismatch of a few percent, this indicates that significant relaxation has taken place. At room temperature, the *ex-situ* application of 6 kV cm[-1] to an as-received unpoled BTO substrate was found to significantly alter the population of 90° domains (Fig. 2b), as expected[17] for fields of 3-10 kV cm[-1] (below the poling field[17] of ~20 kV cm[-1]). Fig. 2b therefore demonstrates that fields of a strength that we can conveniently apply across our 0.5 mm BTO substrates are sufficient to induce dramatic changes in the ferroelectric domains.

ME measurements of Samples #3-5 at various temperatures were performed by monitoring the in-plane pseudo-cubic [100] magnetic response to a voltage applied perpendicularly between the film surface and the substrate underside (Fig. 1, inset). Similar results were obtained in all cases (Fig. 3 and Supplementary Information), as we now discuss. On ramping the voltage, there was initially no significant change in film magnetization. Our main result is the subsequent observation of a large, sharp, persistent transition at some threshold voltage. The associated threshold field did not vary strongly with temperature, just like the coercive field[25] of BTO, but sample to sample variations are likely due to the uncontrolled nature of the BTO domain configurations. Additional increases (Sample #3, Run #3, Supplementary Information) or reversals (Sample #5, Run #2, Supplementary Information) in electric field did not further change the magnetization, consistent with the fact that in BTO the strain is almost fully saturated just above the coercive field[17]. Equally, removing the electric field had no discernible effect on the switched magnetization (Fig. 3), even after ~10 minutes (Sample #5, Run #6, Supplementary Information), consistent with irreversible strain-electric field hysteresis seen in BTO[17].

We were unable to obtain a complete dataset for any given sample because of degradation associated with sample blackening, contact loosening, sparking at high voltages and ultimately cracking (but note that magnetic measurements performed at 70 K and 8000 Oe during poling indicated that previous ME measurements had not damaged film integrity prior to catastrophic failure). Indeed, our experimental conditions were extreme given that ME switching required the applications of hundreds of volts across our macroscopic BTO substrates. The magnitudes of the transitions that we measured (Fig. 4a) show considerable spread, even when a given sample was remeasured at the same temperature. This spread is particularly prounounced near the R-O transition around 200 K, likely due to enhanced domain mobility[22] and a small variability in the set temperature. It is at this temperature that we record our largest value of $\Delta M/M_0$=65% (Fig. 4a). For a single switching event (199 K, Fig. 3), on our crude measurement timescale of 1 s, this value corresponds to our largest $\Delta M$=0.45 $\mu_B$/Mn and our largest $\alpha=\mu_0\Delta M/\Delta E$=2.26×10$^{-7}$ s m$^{-1}$ given $\Delta E$=4 kV cm$^{-1}$. With less clean switching (157 K, Fig. 3), we achieved our largest $\Delta M$=0.69 $\mu_B$/Mn and our largest $\alpha=\mu_0\Delta M/\Delta E$=2.31×10$^{-7}$ s m$^{-1}$ (Fig. 4b), which is a negligible improvement on the above figure in view of the increased $\Delta E$=6 kV cm$^{-1}$. Multistep switching is likely a trivial consequence of our simple experimental procedures, e.g. because the top silver dag contact does not completely cover the film (Fig. 1, inset). However, we note that the switching was sharp in a run where the electric field was turned on abruptly (Sample #3, Run #2, Supplementary Information) rather than ramped.

These values of $\alpha\sim10^{-7}$ s m$^{-1}$ recorded in our epitaxial LSMO//BTO structures significantly exceed the values obtained in single phase materials, whether they arise in the form $dP/dH$ (e.g. 4.1×10$^{-12}$ s m$^{-1}$ in Cr$_2$O$_3$)$^2$ or $\Delta P/\Delta H$ (e.g. ~10$^{-9}$ s m$^{-1}$ in Ni$_3$B$_7$O$_{13}$I and Tb$_2$(MoO$_4$)$_3$)$^{3,14}$. Our results also represent an order of magnitude improvement over values associated with two phase systems, namely the previously discussed estimate of $dP/dH$ (~10$^{-8}$ s m$^{-1}$ in laminated PZT/terfenol-D [5]), and also the recent report[8] of $\mu_0\Delta M/\Delta E$ for CoFe$_2$O$_4$ nanopillars in a BiFeO$_3$ matrix (~10$^{-8}$ s m$^{-1}$ with the factor of $\mu_0$ correctly inserted, but a direct reading of the as-presented data gives the even smaller figure of $\mu_0$×(20 emu cm$^{-3}$)/(10 V / 200 nm)=5×10$^{-10}$ s m$^{-1}$, and both of these figures would be further reduced if the magnetization were normalised using the sample volume rather than the volume of the distributed magnetic material). Note that in [8] the switching was deduced by taking data before and after, but not during the application of the field. Therefore the switching was not shown to be sharp, and indeed it might not be sharp given the complex strain fields and the observation that 1 T was required for magnetic saturation.

We now discuss details of the mechanism responsible for our ME coupling, presumed to be strain-mediated (Fig. 2), under the assumption that the LSMO film is unaffected by electric field effects due to BTO switching. The electrically driven switching of ferroelectric non-180° domains in the R, O and T phases of BTO produces in-plane piezoelectric strains that are expected to modify the overlying ferromagnetic LSMO by locally switching its strain. This in turn locally switches the magnetic anisotropy of the film, and thus the direction of the local magnetization, such that the component detected by the magnetometer is altered. (Note that our measurement procedure precludes in-plane angular scans, but transverse in-plane domains are liable to cancel.) Although we applied no external stresses, the ferroelastic nature of BTO is exploited indirectly for two reasons. Firstly, ferroelastics necessarily possess the non-180° domains required for switching strain states in an overlying epitaxial film. Secondly, ferroelastics typically display large piezoelectric effects. By contrast, non-ferroelastic ferroelectrics such as LiNbO$_3$ possess only 180° domains and would not produce the effects described above. In the room temperature T phase, poled BTO would be unattractive because of the difficulty of electrically creating 90° domains[26]. More generally, the presence of ferroelastic domains facilitates ferroelectric switching[23], consistent with finding our largest (large) values of $\Delta M/M_0$ near the R-O (O-T) transition where more (fewer) domain orientations are permitted. Despite the inhomogeneity that arises from the presence of domains, the use of a single interface simplifies the strain mediated ME coupling with respect to more complex geometries[6-8], and thus permits the large sharp switching observed.

The above interpretation permits the possibility of achieving the converse ME effect, viz. $\Delta P/\Delta H$ or $dP/dH$, in samples where the LSMO:BTO volume ratio is not too low as it is here. Again, switching would be facilitated by the presence of ferroelectric domains, which may be created in epitaxial films by selecting appropriate film thicknesses[27] and deposition conditions[28], and in single crystals by the application of compressive stress[29]. Nevertheless, in a fully $c$-oriented PZT film on a $La_{1.2}Sr_{1.8}Mn_2O_7$ substrate, an effect[30] of estimated strength $dE/dH$=600 mV/ cm Oe (i.e. $dP/dH$=$10^{-9}$ s $m^{-1}$ using[15] $\varepsilon$=300 as before) arose, but only near the ferromagnetic transition at 105 K.

At a single and clean epitaxial interface, we have demonstrated a large, sharp, and persistent ME switching in small electric fields over a wide range of temperatures including room temperature. This advance arises because the homogeneity of the coupling strain is increased with respect to more complex structures[6-8]. Strain components perpendicular to the planar interface are minimised, and in-plane homogeneity is able to extend over distances bounded by the size of the ferroelectric domains. The sharpness of the ME transitions is ultimately limited by the switching speed of these ferroelectric domains, which is approximately eight orders of magnitude faster than the crude 1 s time resolution here.

Several improvements to the present work are apparent. A surface coating of oil would eliminate sparking; a reduction in BTO thickness would both reduce the tendency for cracking and permit the application of larger fields with smaller voltages (e.g. 0.1 V across 200 nm); unpoled and therefore cheaper ferroelectric materials would be desirable given the role of domains; and reversible switching for memory devices[31] could be achieved using defects obtained by ageing[32] or the application of stress[33]. Regarding materials selection, there is scope for increasing $\alpha$ using the present choice of LSMO and BTO for two reasons. Firstly, our values of $\Delta M$ exceed the thermally induced zero-field magnetic jumps both here (Fig. 1) and in Lee $et$ $al.$[16]. Secondly, the values of $\Delta E$ required for switching could be reduced for optimal ferroelectric domains configurations, achieved e.g. at the coercive field on a major hysteresis loop which requires fields of the poling strength. By exploring alternative materials, one could exploit non-hysteretic relaxor ferroelectrics where the piezoelectric response is large and cracking is absent; and larger magnetostrictive or piezomagnetic effects in magnetic materials that are more sensitive to stress.

**Acknowledgements**
We thank M. Cain, G. Catalan, C.-B. Eom, I. C. Infante, E. C. Israel and F. D. Morrison for useful discussions. W. E. is grateful to the EU for a Marie Curie Fellowship. Support from the UK EPSRC and The Royal Society is gratefully acknowledged.

**Corresponding author NDM:** ndm12@cam.ac.uk

**Methods**
40 nm LSMO films were grown by pulsed laser deposition (KrF laser, $\lambda$= 248 nm) on one side polished 4 mm × 4 mm × 0.5 mm $BaTiO_3$ (001) substrates at 775 °C. During deposition, the flowing oxygen pressure was 15 Pa, the laser fluence was 1.5 J/cm$^2$, and the target-substrate distance was 8 cm. After deposition, films were cooled at 5 °C/minute to 700 °C, annealed for 30 minutes at 700 °C in 0.5 atm. $O_2$, and cooled to room temperature at 10 °C/minute. The lattice parameters and film crystallinity were studied by x-ray diffraction (XRD) using a Philips X'Pert high resolution diffractometer ($\lambda$= 1.54 Å) with a Ge monochrometer. Film surfaces were studied with a Digital Instruments Nanoscope III atomic force microscope (AFM) operated in tapping mode. The surface was smooth with a roughness of ~1.0 nm away from twin boundaries. Magnetic properties were investigated in a Princeton Measurements Corporation vibrating sample magnetometer (VSM), with the magnetometer axis ∥ pseudo-cubic [100]. Electric fields were applied in the VSM using a bespoke probe constructed in-house. It was built to the design of Princeton Measurements Corporation, but included electrical wiring. Voltages applied to the bare substrate (Fig. 2b) were applied via sticky copper strips, with additional silver dag on the unpolished side. Voltages applied for ME measurements (Fig. 3) were applied via silver dag contacts on the substrate underside and most but not all of the film surface (see e.g. Fig. 1, inset).

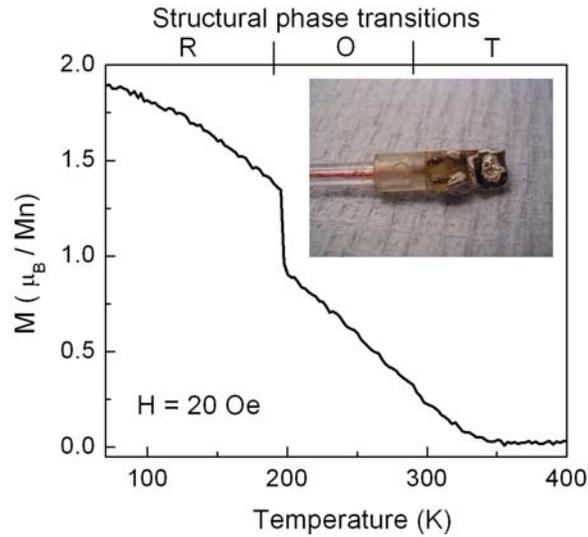

**Figure 1 | Response of a ferromagnetic epitaxial LSMO film to structural phase transitions in the BTO substrate.** The film magnetisation *M* measured on warming is shown for a representative sample (Sample #1) of LSMO(40 nm)//BTO(0.5 mm) measured in 20 Oe, after pre-cooling from 470 K to 70 K in 8000 Oe. The jump of 0.45 $\mu_B$/Mn near 200 K is associated with the R→O structural phase transition in the BTO substrate. In one unrepresentative sample (Sample #2, Supplementary Information), this jump was smaller (0.32 $\mu_B$/Mn) and a second jump (0.17 $\mu_B$/Mn) near 300 K associated with the O→T transition was observed. **Inset,** the VSM probe head with the sample (black) connected via silver dag (grey) and copper leads to wires (red) running to the top of the probe.

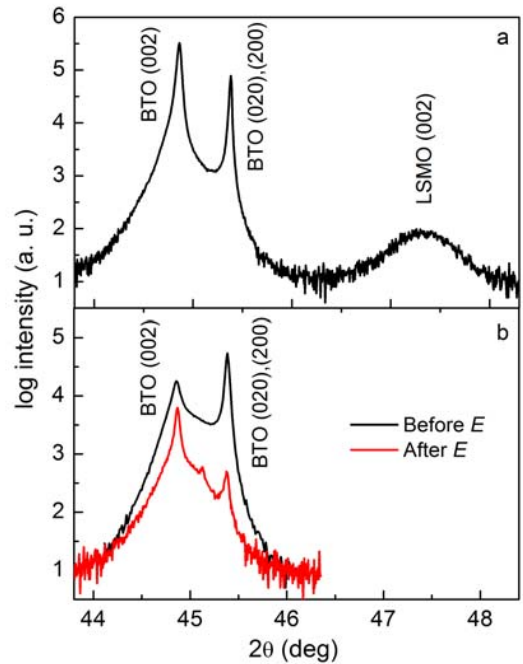

**Figure 2 | Ferroelectric domains in BTO. a,** XRD ω-2θ scan for an as-prepared sample (Sample #1) of LSMO//BTO showing 90° domains in BTO, and the LSMO film peak. **b,** XRD ω-2θ scan for an unpoled BTO substrate before and after the application of $E$=6 kV cm$^{-1}$ between the top and bottom surfaces. The domain population is significantly modified by the field, but the change cannot be formally quantified given the large background signal. However, we note that the ratio of peak heights changes dramatically from 0.3:1 to 12:1. The small additional peak at 45.13° corresponds closely to the $a \approx c$ lattice parameters of orthorhombic twin variants[34].

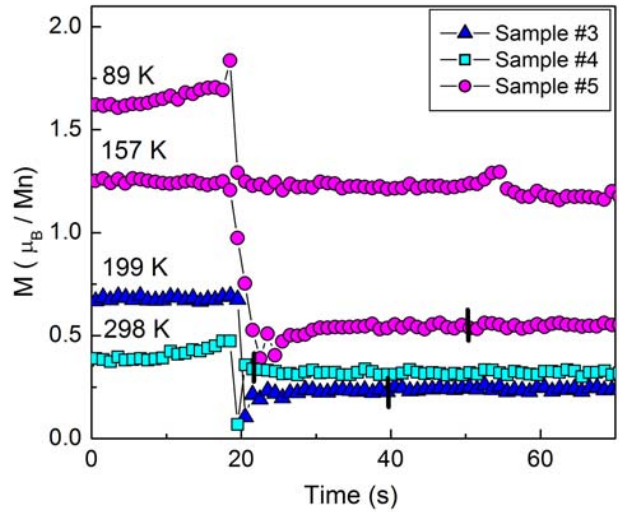

**Figure 3 | Large sharp magnetic switching due to an applied electric field.** Each data sweep was performed in 20 Oe, after an excursion in 8000 Oe from 470 K, to 70 K, to the measurement temperature. The electric field was manually ramped over a period of ~10 s while visually monitoring the VSM output in real time. Once a magnetic transition was observed, the ramping was halted as fast as the operator could respond in order to reduce the chances of breakdown. The switching transient at 298 K is attributed to an observed sparking. The magnetic transitions shown have all been arbitrarily placed near 20 s for clarity. The switching was found to persist after the electric field was abruptly switched off (vertical black lines). (Films biased positive relative to substrates, but reverse bias used for Sample #5 at 157 K. Transitions recorded at 4 kV cm$^{-1}$ (Sample #3), 10 kV cm$^{-1}$ (Sample #4) and 6 kV cm$^{-1}$ (Sample #5 at both 89 K and 157 K).)

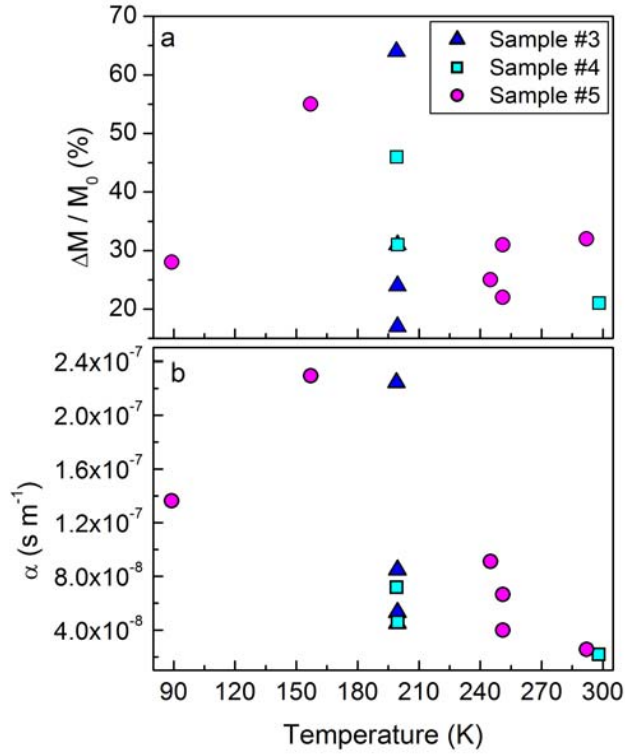

**Figure 4 | Electrically induced magnetic switching and magnetoelectric coupling strength. a,** the jump $\Delta M$ between stable states as a percentage of the zero field starting value $M_0$, for the data presented in Fig. 3 and similar data (Supplementary Information) collected on the same three samples prior to experimental failure. The pronounced spread near 200 K is likely due to the proximity of the R-O phase transition. **b,** the corresponding magnetoelectric coupling $\alpha = \mu_0 \Delta M / \Delta E$ also depends on the electric field $\Delta E$ required for switching the uncontrolled ferroelectric domain configurations of the BTO substrates.

# Supplementary Information

## Giant sharp magnetoelectric switching in multiferroic epitaxial La$_{0.67}$Sr$_{0.33}$MnO$_3$ on BaTiO$_3$

### W. Eerenstein, M. Wiora, J.L. Prieto, J.F. Scott and N.D. Mathur

We present here magnetization versus temperature plots for all five LSMO//BTO samples, including the data for Sample #1 that was shown in Fig. 1 of our Letter. We also present all magnetoelectric measurements made for Samples #3-5, including the data shown in Fig. 3 of our Letter.

### 1) Magnetization versus temperature for all five samples

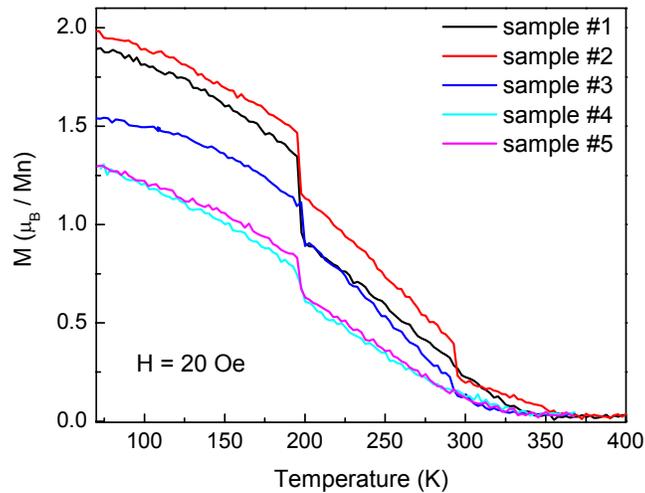

**Figure S1 | Response of ferromagnetic LSMO epitaxial films to structural phase transitions in BTO substrates (Samples #1-5).** Magnetisation $M$ measured on warming in 20 Oe, after pre-cooling from 470 K to 70 K in 8000 Oe. The jump near 300 K associated with the O→T transition was only unambiguously observed in Sample #2. The other four samples behave in a very similar manner.

## 2) Magnetoelectric dataset (Samples #3-5)

Here we summarise and present all of the magnetoelectric data that we collected by measuring the magnetometer output when applying an electric field. As explained in Fig. 3 of our Letter, each data sweep was performed in 20 Oe, after an excursion in 8000 Oe from 470 K, to 70 K, to the measurement temperature. Subsequently, the electric field was manually ramped over a period of ~10 s while visually monitoring the VSM output in real time. Once a magnetic transition was observed, the ramping was halted as fast as the operator could respond in order to reduce the chances of breakdown. Films were biased positive relative to substrates. Two exceptions to the above protocol are listed below and noted in the table in a):

- Sample #3, Run #2: The electric field was turned on suddenly rather than ramped (and a sharp transition was observed).
- Sample #5, Run #2: The electric field was applied in the opposite sense (producing the largest values of $\Delta M$ and $\alpha$, but perhaps not as a consequence of this sign reversal).

### a) Magnetoelectric data summary

The following table summarises the data for all magnetoelectric switching events that we observed. Entries in boldface type correspond to the data presented in Fig. 3 of our Letter. All of the raw data is presented in the following sub-section.

| Sample-Run | T (K) | $M_0$ ($\mu_B$/Mn) | $\Delta M$ ($\mu_B$/Mn) | $\Delta M/M_0$ (%) | $\Delta E$ (kV cm$^{-1}$) | $\alpha = \mu_0 \Delta M / \Delta E$ (s m$^{-1}$) | $(\mu_0 \Delta M/M_0) / \Delta E$ (s A$^{-1}$) | Single switch? | Notes |
|---|---|---|---|---|---|---|---|---|---|
| **3-1** | **199** | **0.68** | **0.45** | **0.66** | **4** | **2.26×10$^{-7}$** | **3.3×10$^{-7}$** | **Y** | |
| 3-2 | 199.5 | 0.6 | 0.09 | 0.15 | 4 | 4.52×10$^{-8}$ | 7.5×10$^{-8}$ | Y | $\Delta E$ turned on suddenly with no ramp |
| 3-3 | 199.5 | 0.52 | 0.16 | 0.31 | 4 | 8.04×10$^{-8}$ | 1.5×10$^{-7}$ | N | No increase in $\Delta M$ with additional $\Delta E$ |
| 3-4 | 199.5 | 0.75 | 0.17 | 0.23 | 4 | 8.54×10$^{-8}$ | 1.1×10$^{-7}$ | N | |
| **4-1** | **298** | **0.42** | **0.11** | **0.26** | **10** | **2.21×10$^{-8}$** | **5.3×10$^{-8}$** | **N** | |
| 4-2 | 199.1 | 0.77 | 0.36 | 0.47 | 10 | 7.24×10$^{-8}$ | 9.4×10$^{-8}$ | N | |
| 4-3 | 199.6 | 0.77 | 0.23 | 0.30 | 10 | 4.62×10$^{-8}$ | 6.0×10$^{-8}$ | N | |
| **5-1** | **89** | **1.63** | **0.41** | **0.25** | **6** | **1.37×10$^{-7}$** | **8.4×10$^{-8}$** | **Y** | |
| **5-2** | **157** | **1.26** | **0.69** | **0.55** | **6** | **2.31×10$^{-7}$** | **1.8×10$^{-7}$** | **N** | **$\Delta E$ applied with opposite sign** |
| 5-3 | 251 | 0.63 | 0.2 | 0.32 | 6 | 6.70×10$^{-8}$ | 1.1×10$^{-7}$ | N | |
| 5-4 | 251 | 0.53 | 0.12 | 0.23 | 6 | 4.02×10$^{-8}$ | 7.6×10$^{-8}$ | N | |
| 5-5 | 292 | 0.26 | 0.09 | 0.35 | 7 | 2.58×10$^{-8}$ | 9.9×10$^{-8}$ | N | |
| 5-6 | 245 | 1.3 | 0.32 | 0.25 | 7 | 9.19×10$^{-8}$ | 7.1×10$^{-8}$ | Y | Reversal of $E$ has no effect. Persistent switch observed during ~10 minutes |

**Table S1 | ME data summary.** All values of $\Delta M/M_0$ are plotted in Fig. 4a of our Letter, and all values of $\alpha$ appear in Fig. 4b.

**b) Raw magnetoelectric data**

Here we show all of the magnetization versus time plots that we acquired. Each plot corresponds to an entry in the table on the previous page. Four of the plots appear in Fig. 3 of our Letter but are reproduced here for completeness. As in Fig. 3, vertical black lines indicate when the electric field was abruptly switched off. In some runs (e.g. Sample #4, Run #1) there are switching transients, which we attribute to an observed sparking. When this happened the electric field was switched off as quickly as possible.

**Sample #3, Run #1**

These data appear in Fig. 3 of our Letter.

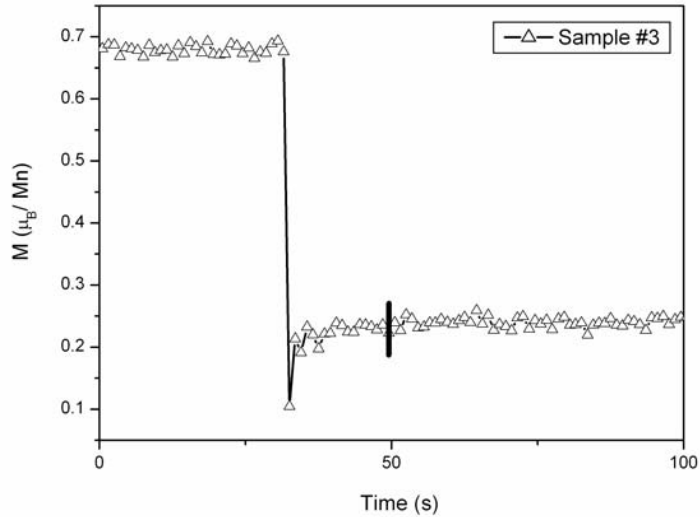

**Sample #3, Run #2**

The field was switched on by applying 200 V suddenly at 50 s, without ramping.

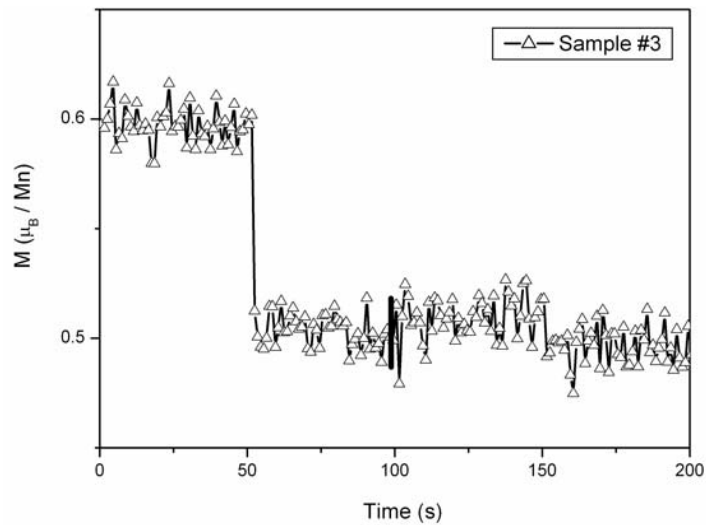

**Sample #3, Run #3**

4 kV cm$^{-1}$ was applied suddenly without ramping, and then increased to the threshold field of 6 kV cm$^{-1}$. This resulted in sparking and so the field was removed within seconds.

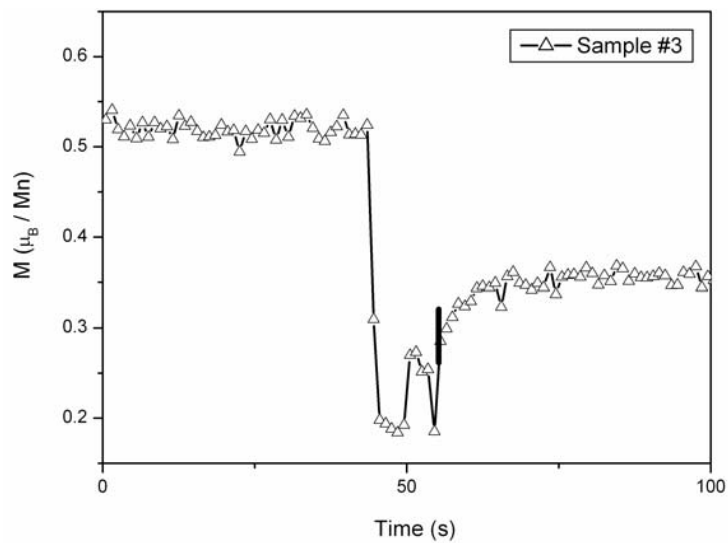

**Sample #3, Run #4**

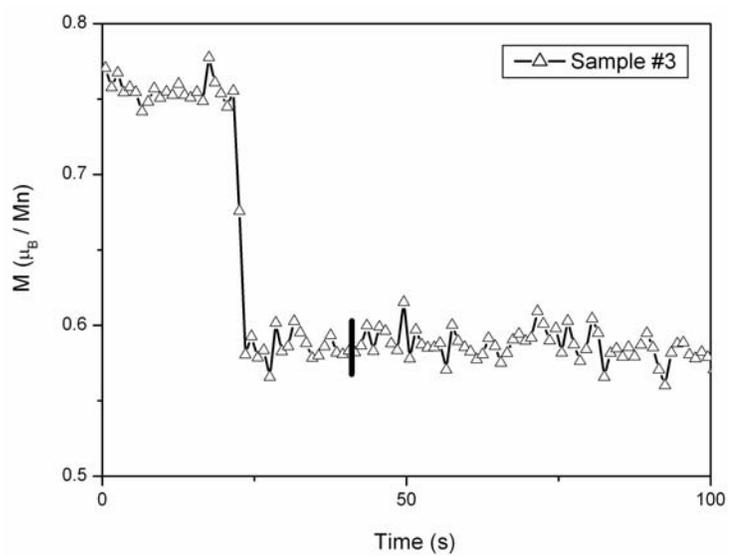

**Sample #4, Run #1**
Sparking seen after the transition, and so the field was removed within seconds. These data appear in Fig. 3 of our Letter.

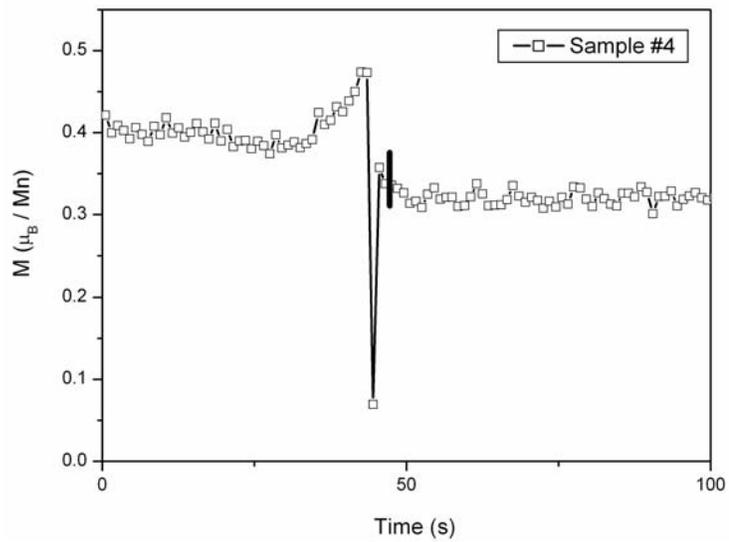

**Sample #4, Run #2**
Sparking seen after the transition, and so the field was removed within seconds.

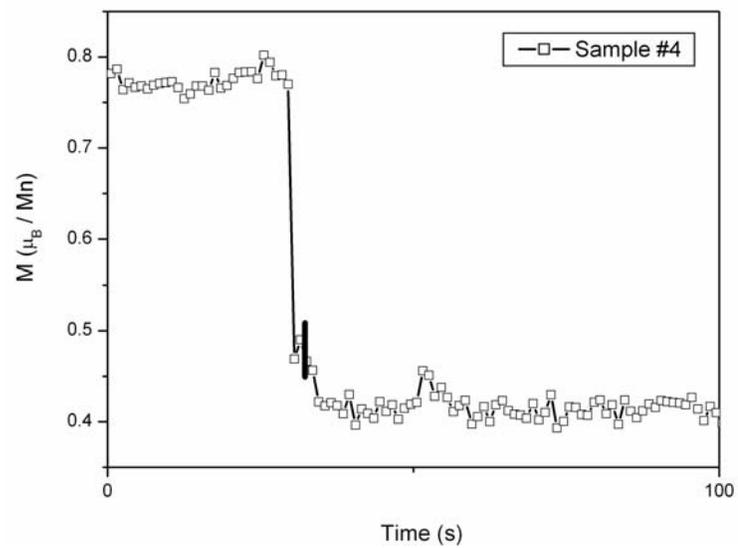

**Sample #4, Run #3**
Sparking seen after the transition, and so the field was removed within seconds.

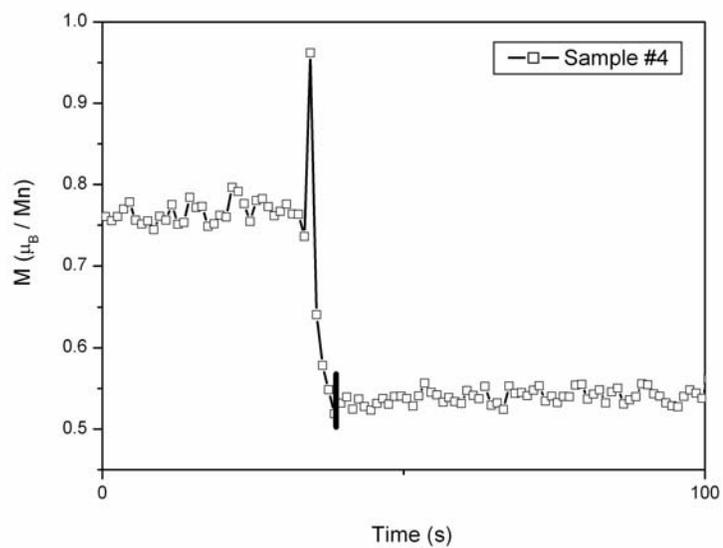

**Sample #5, Run #1**
Electric field left on for undefined length of time. These data appear in Fig. 3 of our Letter.

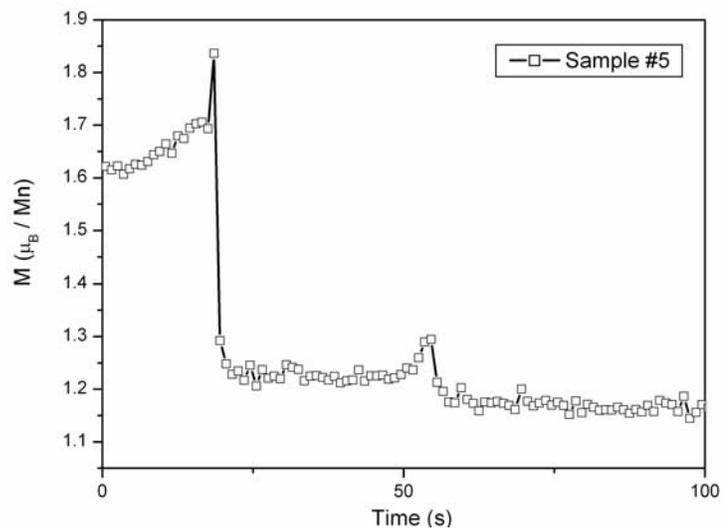

**Sample #5, Run #2**
Electric field applied in opposite sense to all other data sweeps. These data appear in Fig. 3 of our Letter.

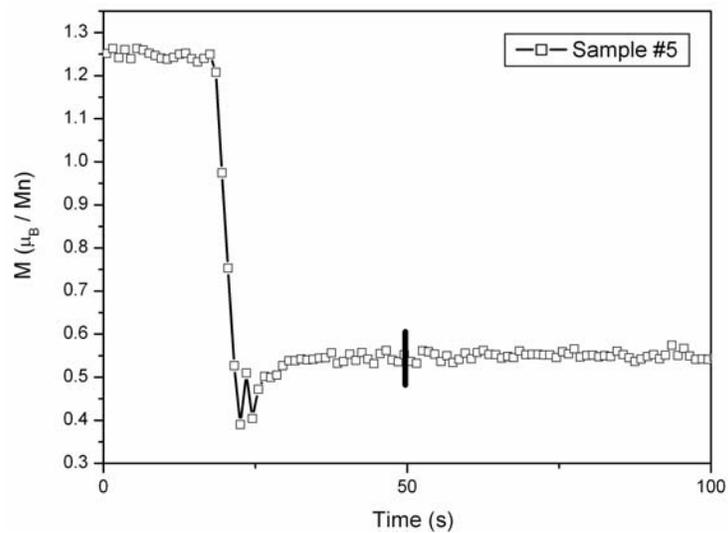

**Sample #5, Run #3**
Electric field left on for undefined length of time.

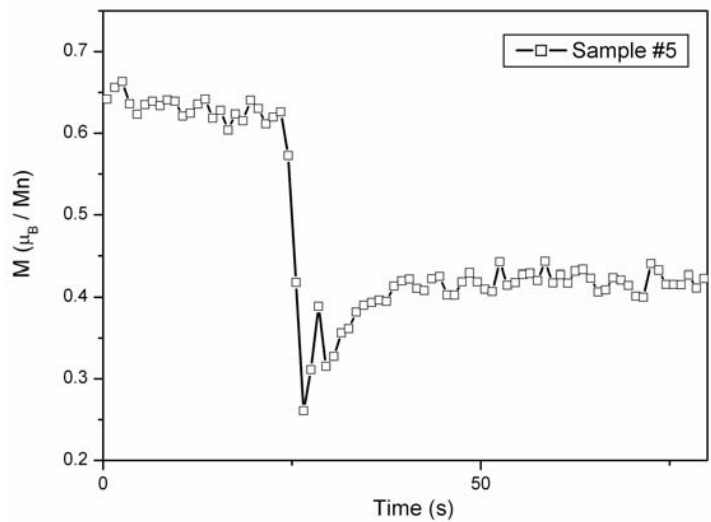

**Sample #5, Run #4**

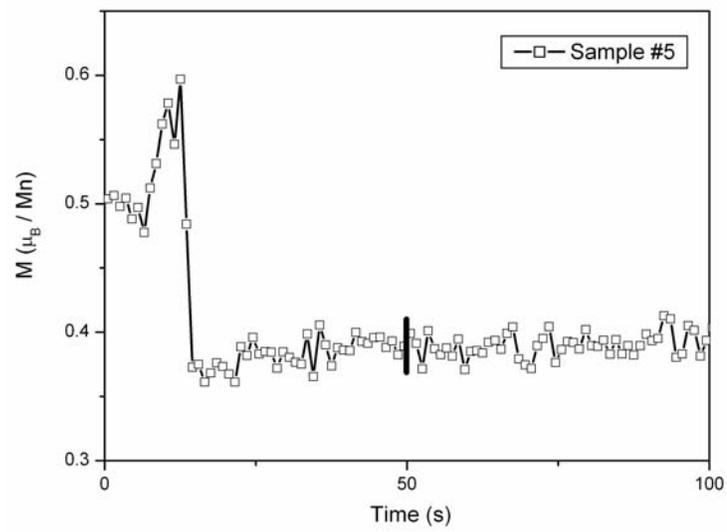

**Sample #5, Run #5**

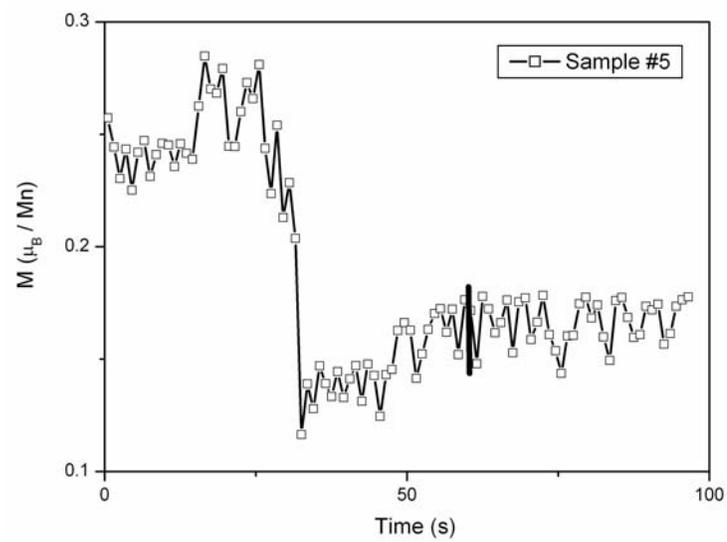

**Sample #5, Run #6**

A reverse field of the switching strength was applied suddenly at 500 s, but no effect was observed. This reverse field was removed at an undefined time. Note that the observed switch was persistent during the ~10 minute monitoring period.

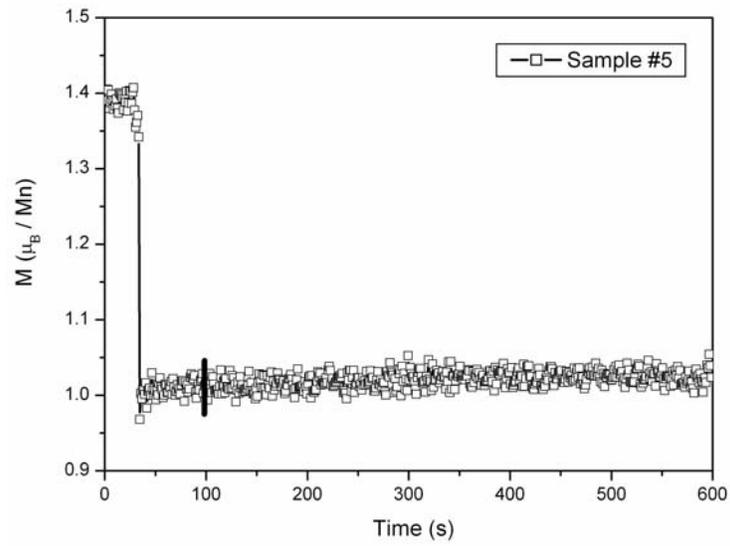